\documentclass[aip,jcp,reprint,citeautoscript,floatfix]{revtex4-1}
\usepackage{natbib}
\usepackage{graphicx}
\usepackage[caption=false,lofdepth,lotdepth]{subfig}
\usepackage{import}
\usepackage{lmodern}
\usepackage[colorinlistoftodos,color=black!30]{todonotes}
\usepackage{amsmath, amsthm, amssymb, mathtools}
\usepackage[T1]{fontenc}
\usepackage{microtype}
\usepackage[utf8]{inputenc}
\usepackage{url}
\usepackage{enumerate}
\usepackage{physics} 
\usepackage{bm} 
\usepackage{booktabs,tabularx}
\usepackage{cleveref} 
\usepackage{accents}
\usepackage{layouts}
\usepackage{booktabs}
\usepackage{float}
\usepackage{placeins}
\usepackage{tikz-cd}

\usepackage{algorithm}
\usepackage{algpseudocode}
\usepackage{caption}
\crefname{equation}{Eq.}{Eqs.}
\crefname{section}{Sec.}{Secs.}
\crefname{table}{Tab.}{Tabs.}
\crefname{figure}{Fig.}{Figs.}
\crefname{subfigure}{Fig.}{Figs.}

\usepackage[nolist,nohyperlinks]{acronym}
\usepackage{bm}
\usepackage{pgfplots}
\usetikzlibrary{pgfplots.groupplots}

\pgfplotsset{compat=1.5}

\usepackage{siunitx} 
\usepackage{multirow} 
\usepackage{bbm}

\usepackage{algorithm}
\usepackage{algpseudocode}

\usepackage{enumitem}

\graphicspath{{figures/}}

\begin{document}
\definecolor{plt_blue}{RGB}{31,119,180}
\definecolor{plt_orange}{RGB}{255,127,14}
\definecolor{plt_green}{RGB}{44,160,44}
\definecolor{plt_red}{RGB}{214,39,40}

\title{Multi-Resolution  Wire-Fencing for Efficient Path Sampling} 
\author{Simen Z. Stenersen Michler}
\affiliation{Norwegian University of Science and Technology, Department of Chemistry and Biomedical Science,  NO-7491 Trondheim, Norway}
\author{Lukas Baldauf}
\affiliation{Norwegian University of Science and Technology, Department of Chemistry and Biomedical Science,  NO-7491 Trondheim, Norway}
\author{Titus S. van Erp}
\affiliation{Norwegian University of Science and Technology, Department of Chemistry and Biomedical Science,  NO-7491 Trondheim, Norway}

\begin{abstract}
Path sampling methods enable the computation of thermodynamic and kinetic properties through Monte Carlo (MC) moves that generate trajectories from short forward and backward molecular dynamics (MD) segments. Recently, the  wire-fencing  move was developed to achieve near-unity acceptance while also rapidly decorrelating successive paths, two properties that are usually in conflict in conventional MC schemes. However, in large systems, such as biomolecular simulations, the frame-saving frequency is often kept low to reduce storage requirements. Likewise, when evaluating the order parameter is expensive, frames are saved less frequently to reduce the associated cost. In either situation, this can severely limit the number of available shooting points, in extreme situations leaving only a single point, typically the one with the highest order parameter value, accessible for shooting. Repeated shootings may then originate from the same configuration, reducing sampling efficiency. Here, we introduce a multi-resolution variant of the  wire-fencing  move in which selected subtrajectories are propagated at higher temporal resolution than the stored trajectories. This refinement affects only the MC move and does not alter the structure or storage of the generated paths, but enhances the diffusion of shooting points along the trajectory and thereby improves sampling efficiency. 
The approach is demonstrated on two model systems and a realistic protein–ligand unbinding process, 
with the latter showing an estimated efficiency improvement of more than an order of magnitude.
\end{abstract}
\maketitle
\section{Introduction}
\label{sec:introduction}
Rare events, such as chemical reactions, phase transitions, permeation, protein folding, or ligand unbinding, pose a major challenge for brute-force molecular dynamics (MD), as the waiting time before an event occurs can be orders of magnitude longer than the accessible MD time scale.\cite{Peters2017}
 Many rare-event methods address this by modifying the underlying potential energy surface or dynamics, which allows exact thermodynamic calculations but sacrifices information about the spontaneous dynamics. Path sampling approaches\cite{TPS, TPSReview2, TIS, RETIS,Hall2022RETISFFS,vanErp2023RETIS}, in contrast, collect short MD trajectories via a
Monte Carlo (MC) procedure, providing exact kinetic and thermodynamic information without altering the dynamics.
The original transition path sampling (TPS)\cite{TPS} algorithm primarily focused on the exploration of transition pathways, but also provided a framework for rate calculations, which was later improved by transition interface sampling (TIS)\cite{TIS}. Subsequent methodological developments, including replica-exchange TIS (RETIS) and more recently $\infty$RETIS,\cite{InfRET1, PNAS2024} have further enhanced both efficiency and accuracy, enabling the exact computation of rate constants and equilibrium properties.

Path sampling methods exploit the fact that, while rare events have long waiting times, once a transition initiates it typically proceeds rapidly. They bypass the long waiting times by focusing on barrier crossings and crossing attempts. In particular, TIS methods describe different stages of the reaction through a series of path ensemble simulations, each defined by a minimal progress condition corresponding to the crossing of an interface, which is specified as a fixed value of the order parameter. The overall small probability of the reaction is then obtained as a product of conditional probabilities from these ensembles, each quantifying the likelihood of reaching the next interface given that the previous one has already been crossed.

The main difference between TIS and RETIS is that the latter incorporates replica-exchange moves between path ensembles, which substantially improves CPU efficiency. However, RETIS is more difficult to parallelize because paths have different lengths, making replica synchronization challenging. This limitation is overcome in $\infty$RETIS, which enables massive parallel execution by employing asynchronous replica exchange with effectively infinite swaps. In addition, the main MC move for generating new paths within an ensemble has been improved through subtrajectory shooting.\cite{riccardi2017fast} Among these moves,  wire-fencing \cite{WF}  has become the default choice in $\infty$RETIS due to its simplicity and efficiency, achieving near-unity acceptance while maintaining fast decorrelation.

Despite these advances, several challenges remain\cite{vanErp2023RETIS}. In systems where events are both rare and slow, exact path sampling methods are no longer applicable, and alternative approaches such as partial path TIS (PPTIS),\cite{moroni2004rate} Milestoning,\cite{faradjian2004computing} or replica exchange PPTIS (REPPTIS)\cite{vervust2023path} are typically employed. However, these methods rely on Markovian assumptions, which require a careful choice of order parameter and interface placement; otherwise, the assumptions may break down and lead to inaccurate results. Non-equilibrium systems pose additional challenges, as backward-in-time propagation is often not possible and the steady-state probability distribution is generally unknown. Splitting-based methods such as forward flux sampling (FFS),\cite{Hall2022RETISFFS} weighted ensemble (WE)\cite{huber1996weighted} dynamics, or adaptive multilevel splitting (AMS)\cite{cerou2011multiple} can be applied in such cases. However, the absence of backward propagation introduces a significant risk that simulations fail to converge to the most likely mechanism and may strongly underestimate rates by favoring unfavorable pathways.\cite{vanErp2012}

In this work, we address a different challenge that is particularly relevant for large biomolecular systems, as well as for systems in which evaluating the order parameter is computationally expensive, such as nucleation processes. In such systems, long trajectories, large system sizes, and/or expensive order parameter evaluations often enforce a low frame-saving frequency, defined by the subcycle
parameter. For example, a subcycle of 1000 implies that each MD call propagates at least 1000 steps before the path sampling software collects a configuration, evaluates order parameters, and decides whether to continue or terminate the trajectory.

While a large subcycle reduces storage demands and computational overhead, it also limits the number of available shooting points, particularly on steep barrier slopes. In  wire-fencing, shooting points from the previous trajectory are selected beyond the ensemble interface. In the worst case, this may leave only a single valid shooting point. If MD propagation causes the order parameter to drop below the interface within one subcycle in both time directions, subsequent trajectories or subtrajectories can only be launched from the same configuration. Standard shooting offers more flexibility in selecting shooting points, but typically performs worse, as choosing points below the interface often generates paths that fail to cross it, leading to rejected trajectories after significant MD effort. Wire-fencing partially mitigates this issue by generating multiple short subtrajectories (also called subpaths) before extending the final one into a full path, thereby increasing the chance of advancing the shooting point. However, many of these subtrajectories, although short, still do not contribute effectively to exploring path space.

For this reason, we introduce a multi-resolution  wire-fencing  (MRWF) move, in which the intermediate subtrajectories are propagated at higher temporal resolution, or equivalently at a lower subcycle. This increases the number of available shooting points, which can subsequently seed new subtrajectory generations. High- and low-resolution subtrajectories alternate, with any intermediate rejection falling back to a previous high- or low-resolution subpath. Each MRWF move concludes with a low-resolution subpath, which is then extended to a full trajectory, ensuring that all generated paths retain the original resolution.

We demonstrate MRWF on a one-dimensional double-well potential and a more complex puckering example, confirming that rates and crossing probabilities match established 
results. 
Finally, we apply MRWF to a biologically relevant protein–ligand unbinding system with a subcycle of 1000, demonstrating clear efficiency gains.

\section{Theory}
\label{Sec:theory}
\subsection{Super-detailed balance and high-acceptance}

In this section, we do not aim to provide a full exposition of the $\infty$RETIS theoretical framework, which is described extensively elsewhere.\cite{InfRET1, PNAS2024} Instead, we focus on the MC path generation move within the $[i^+]$ path ensemble via a high-acceptance super-detailed balance scheme. This scheme forms the basis of all subtrajectory moves, including  wire-fencing  and MRWF.
 To clarify the concept of a path ensemble, we briefly outline the definition of $[i^+]$.

We assume a reaction coordinate, or order parameter, $\lambda(x)$ defined on a phase point $x$, which describes the progress from reactant state $A$ to product state $B$, with boundaries $\lambda_A < \lambda_B$. The transition rate from $A$ to $B$ is then expressed as the flux of trajectories crossing $\lambda_A$ from within $A$, multiplied by the probability that such trajectories reach $\lambda_B$ without recrossing $\lambda_A$. Since this probability is typically very small, path ensembles are introduced to obtain improved statistics on both reactive trajectories and excursions that make substantial progress but ultimately might return to $A$.

In path sampling, a path, or trajectory, of length $L$ is represented as a sequence of time slices $X = \{x_0, x_1, \ldots, x_L\}$, where $x_\ell$ denotes the phase point of the system at time $t = \ell \Delta t$, starting from an initial condition $x_0$. Here, $\Delta t$ is a small time step, typically corresponding to a few MD steps as determined by the subcycle number. For a trajectory to be a valid path in the $[i^+]$ ensemble, it must satisfy $\lambda(x_0) < \lambda_A$ and terminate upon first reaching either $A$ or $B$, such that $\lambda(x_L) < \lambda_A$ or $\lambda(x_L) > \lambda_B$, while for all intermediate time slices $0 < \ell < L$ one has $\lambda_A < \lambda(x_\ell) < \lambda_B$. In addition, the path must exhibit at least one time slice $x_k$ such that $\lambda(x_k) \ge \lambda_i$.

The main goal of path generation moves, such as standard shooting\cite{shoot} and  wire-fencing \cite{WF}, is to sample trajectories for the different path ensembles such that the resulting distribution corresponds to the same statistical distribution as would be obtained by extracting the relevant trajectory segments from an infinitely long MD trajectory.
To achieve this, MC moves are typically designed to obey detailed balance, even though the weaker condition of balance would, strictly speaking, be sufficient. While detailed balance is more restrictive than balance, it is considerably easier to construct consistent acceptance rules within this framework. For complex moves, it can even be beneficial to impose an even stronger condition, such as \emph{super-detailed balance}. 
For  wire-fencing  and MRWF moves, the Metropolis--Hastings acceptance rule for generating a new path from an existing one in the $[i^+]$ ensemble, based on super-detailed balance, can be written as:\cite{riccardi2017fast, WF}
\begin{align}
P_{\rm acc} &= {\mathbbm 1}_{[i^+]}\!( X^{(n)})
\times \label{eq:supdet} \\
&\quad \min\!\left[ 1,
\frac{P(X^{(n)})
P_{\rm gen}(X^{(n)} \rightarrow
X^{(o)} \textrm{ via } \overline{\chi})}{
P(X^{(o)})
P_{\rm gen}(X^{(o)} \rightarrow
X^{(n)} \textrm{ via } \chi)}
\right]
\nonumber
\end{align}
where $P_{\rm acc}$ denotes the acceptance probability of the proposed move, $X^{(o)}$ and $X^{(n)}$ represent the old and new paths, respectively, $P(X)$ is the path probability distribution, $P_{\rm gen}(X^{(o)} \rightarrow X^{(n)} \textrm{ via } \chi)$ is the probability to generate the new path from the old path through the exact sequence of stochastic steps (including, e.g., selection and propagation of subtrajectories) that constitute the move $\chi$, and $P_{\rm gen}(X^{(n)} \rightarrow X^{(o)} \textrm{ via } \overline{\chi})$ is the corresponding reverse generation probability. Finally, ${\mathbbm 1}_{[i^+]}( X)$ denotes an indicator function that equals 1 if the path is valid for the ensemble $[i^+]$ and 0 otherwise.

The notation “via $\chi$” and “via $\overline{\chi}$” imposes additional constraints consistent with the super-detailed balance scheme, implying that the generation probabilities in Eq.~\ref{eq:supdet} account for the full sequence of intermediate subtrajectories. 
The reverse generation probability corresponds to producing the old path from the new one through the same sequence of subtrajectories, but in reverse order (except for rejected subtrajectories; see Sect.~\ref{sec:alg}).
Hence, each $\chi$ has a unique inverse $\overline{\chi}$, and all terms in Eq.~\ref{eq:supdet} 
cancel\cite{WF} except the initial shooting-point selection:
\begin{align}
P_{\rm acc} &= {\mathbbm 1}_{[i^+]}\!( X^{(n)})
\times
 \min\!\left[ 1,
\frac{M_i^{(o)}}{ M_i^{(n)}}
\right]
\label{eq:supdet2}
\end{align}
where $M_i^{(o)}$ and $M_i^{(n)}$ denote the number of available shooting points for initiating the first subtrajectory on the old and new paths, respectively.

The main idea behind subtrajectory moves is that subtrajectories are much shorter than full trajectories and are therefore used as intermediates to eventually construct a new trajectory that is significantly decorrelated from the previous one. As a result, although a single MC move is typically more expensive than standard shooting, far fewer full trajectories are required to achieve a given statistical accuracy. If a subtrajectory is unsuccessful (i.e., an intermediate rejection that does not lead to the rejection of the MC move itself), a new one is launched from the most recently accepted subtrajectory. However, if the final subtrajectory is extended to a full trajectory and rejected via Eq.~\ref{eq:supdet2}, the entire MC move is rejected and the next move restarts from the original path.

The high-acceptance scheme mitigates this inefficiency.\cite{riccardi2017fast, WF} If the final extended trajectory starts in $B$ but ends in $A$, it is reversed to yield an $A \rightarrow B$ path, effectively doubling the generation probability to account for initially choosing the wrong time direction. This factor is captured by the path function $q(X)$, with $q(X)=2$ for $A \rightarrow B$ and $q(X)=1$ for $A \rightarrow A$. It is combined with $M_i(X)$ to define the high-acceptance weight $w_i(X) = q(X) M_i(X)$ and the biased distribution $\tilde{P}(X) = P(X) w_i(X)$.
Since the sampling distribution has now changed, applying Eq.~\ref{eq:supdet} to $\tilde{P}(\cdot)$ instead of $P(\cdot)$ leads to the cancellation of both $M_i$ terms and the extra factor of 2 in the generation probabilities, yielding
\begin{align}
P_{\rm acc} &= {\mathbbm 1}_{[i^+]}\!( X^{(n)})
\label{eq:supdet3}
\end{align}
In  wire-fencing, any trial path has at least one phase point $x_k$  satisfying $\lambda(x_k) > \lambda_i$, 
so super-detailed balance requires rejection only if the final path reaches $B$ in both time directions.
In the post-simulation analysis, each sampled path $X$ is weighted by $1/w_i(X)$ so that the results again reflect the correct distribution $P(\cdot)$ rather than the biased distribution $\tilde{P}(\cdot)$.

As we will show, the MRWF algorithm introduced in the next section follows the same principles as standard  wire-fencing, in that for each complex sequence of subtrajectory moves $\chi$ there exists a unique reverse sequence 
$\overline{\chi}$
leading to the exact same cancellation of terms in the Metropolis--Hastings scheme. Therefore, the same acceptance rule, Eq.~\ref{eq:supdet3}, and corresponding unbiasing weight $w_i(X)$ are valid for MRWF as for standard  wire-fencing.

\subsection{Wire-fencing}
\label{sec:alg}
\begin{figure}[hb!]
    \centering
\includegraphics[width=\linewidth]{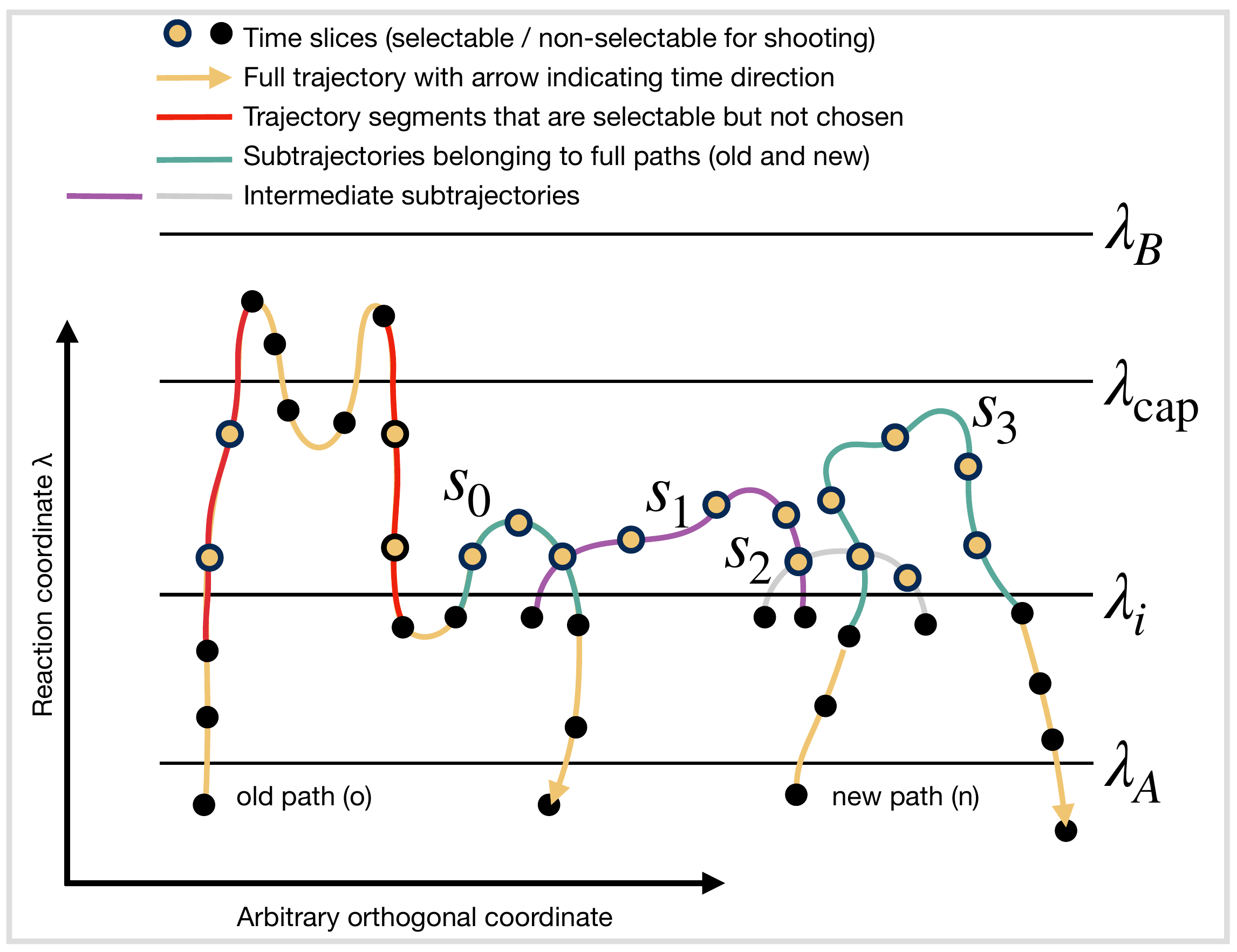}
    \caption{
    Schematic illustration of a   wire-fencing  (WF) move in the $[i^+]$ ensemble, showing how a new path is generated through a sequence of shorter subtrajectories.
    }
    \label{fig:WF}
\end{figure}

Fig.~\ref{fig:WF} illustrates a properly functioning   wire-fencing  (WF) move in the $[i^+]$ ensemble. The relevant interfaces are indicated: the state boundaries $\lambda_A$ and $\lambda_B$, the interface $\lambda_i$ that every path in $[i^+]$ must cross, and the capping interface $\lambda_{\rm cap}$. The latter is introduced to limit the lengths of the subtrajectories and to avoid shooting attempts originating deep within the basin of attraction of state $B$.

The left-hand trajectory (yellow) represents the old path, where the circles denote time slices separated by the standard frame-saving interval. Here, a \emph{segment} is defined as the portion of a trajectory bounded by two consecutive crossings of either $\lambda_i$ or $\lambda_{\rm cap}$. The old path contains seven selectable shooting points distributed over three segments according to the wire-fencing selection criterion: a shooting point must lie between $\lambda_i$ and $\lambda_{\rm cap}$, and its segment (shown in red or green) must cross $\lambda_i$ at one or both endpoints.

Since the path contains $M_i(o)=7$ selectable shooting points and terminates at $\lambda_A$, we have $q=1$, giving a high-acceptance path weight of $w_i(o)=q(o)\,M_i(o)=1\times7=7$. One of the seven selectable shooting points is selected, and its corresponding segment (green) defines the zeroth subtrajectory, $s_0$.
The algorithm then performs a series of $N_{\rm subpath}$ successive uniform shooting moves to create new subtrajectories, where $N_{\rm subpath}$ is a user-defined parameter of the wire-fencing algorithm. In the present example, $N_{\rm subpath}=3$. 

During each shooting move, any time slice  (yellow) of the current subtrajectory, except for the endpoints (black), may be selected. The particle velocities are regenerated from a Maxwell--Boltzmann distribution, after which molecular dynamics trajectories are propagated forward and backward in time from the perturbed phase point until either $\lambda_i$ or $\lambda_{\rm cap}$ is reached, thereby generating the next subtrajectory. Finally, the last subtrajectory ($s_3$ in Fig.~\ref{fig:WF}) is extended backward and forward in time until reaching either $\lambda_A$ or $\lambda_B$, to obtain the new full trajectory.
The final new path  shown in Fig.~\ref{fig:WF} has a 
high-acceptance weight $w_i(n)$  of 5, based on $q(n)=1$ and $M_i(n)=5$, where the selectable shooting points are now distributed over a single segment.

In this case, $\chi$ corresponds to the sequence
$s_0, s_1, s_2, s_3$,
while $\overline{\chi}$ corresponds to the reverse generation sequence
$s_3, s_2, s_1, s_0$.
The figure illustrates a situation in which all subtrajectories reach $\lambda_i$ in at least one time direction and are therefore accepted. Each generated subtrajectory subsequently serves as the starting point for generating the next subtrajectory.

If, however, one of the subtrajectories were to terminate at $\lambda_{\rm cap}$ in both time directions, it would be rejected, while the MC move itself would continue from the last accepted subtrajectory. For example, if $s_2$ were rejected, the algorithm would return to the last accepted subtrajectory, $s_1$, from which a new shooting move would be performed to generate $s_3$. Consequently, $s_3$ and $s_2$ will generally not share a common phase point, and it is therefore not possible to generate $s_2$ from $s_3$, and vice versa.

Thus, while $\chi$ represents the ordered sequence of attempted subtrajectories
$s_0, s_1, s_2, s_3$, the reverse sequence $\overline{\chi}$ is not simply the reverse of this ordering. Instead, it corresponds to the sequence
$s_3, s_1, s_2, s_0$.
Nevertheless, there still exist a symmetry, as illustrated schematically below:
\begin{eqnarray*} 
\begin{tikzcd} \chi= s_0 \arrow[r] & s_1 \arrow[r] \arrow[d, shift right=2] & s_3 \\ & s_2 \arrow[u, shift right=2] \end{tikzcd}\\ \begin{tikzcd} \overline{\chi}= s_3 \arrow[r] & s_1 \arrow[r] \arrow[d, shift right=2] & s_0 \\ & s_2 \arrow[u, shift right=2] 
\end{tikzcd} \end{eqnarray*}
These diagrams make explicit that, following the rejection of $s_2$, the Markov chain returns to the previous accepted subpath, $s_1$, before generating the subsequent subtrajectory.

The bottom line is that, for each complex MC move $\chi$ consisting of multiple subpaths, there always exists a unique reverse move $\overline{\chi}$. This allows the acceptance rule to be constructed via Eq.~\ref{eq:supdet} using the super-detailed-balance principle. Since super-detailed balance implies detailed balance, the resulting scheme ultimately guarantees sampling from the desired target distribution: either the physical distribution $P[X]$ in the case of Eq.~\ref{eq:supdet2}, or the biased distribution $\tilde{P}[X]$ in the case of Eq.~\ref{eq:supdet3}. In the latter case, the bias can be removed from path-ensemble averages by weighting each path with the inverse of its high-acceptance weight.

 Although a wire-fencing move is more computationally expensive than a standard shooting move, the relative increase in cost remains modest when the average subtrajectory length is small compared with the full path length.
The improved efficiency relative to standard shooting arises from the fact that the new path is substantially more decorrelated from the old path. Consequently, substantially fewer full trajectories are required to achieve a given level of convergence. 
The magnitude of this efficiency gain is system-dependent, but for sufficiently complex test models, subtrajectory moves with high acceptance rates have been shown to yield an efficiency improvement of approximately a factor of twelve compared with standard shooting~\cite{riccardi2017fast}.

\subsection{Shooting-point scarcity}

\begin{figure}[ht!]
\centering
\includegraphics[width=\linewidth]{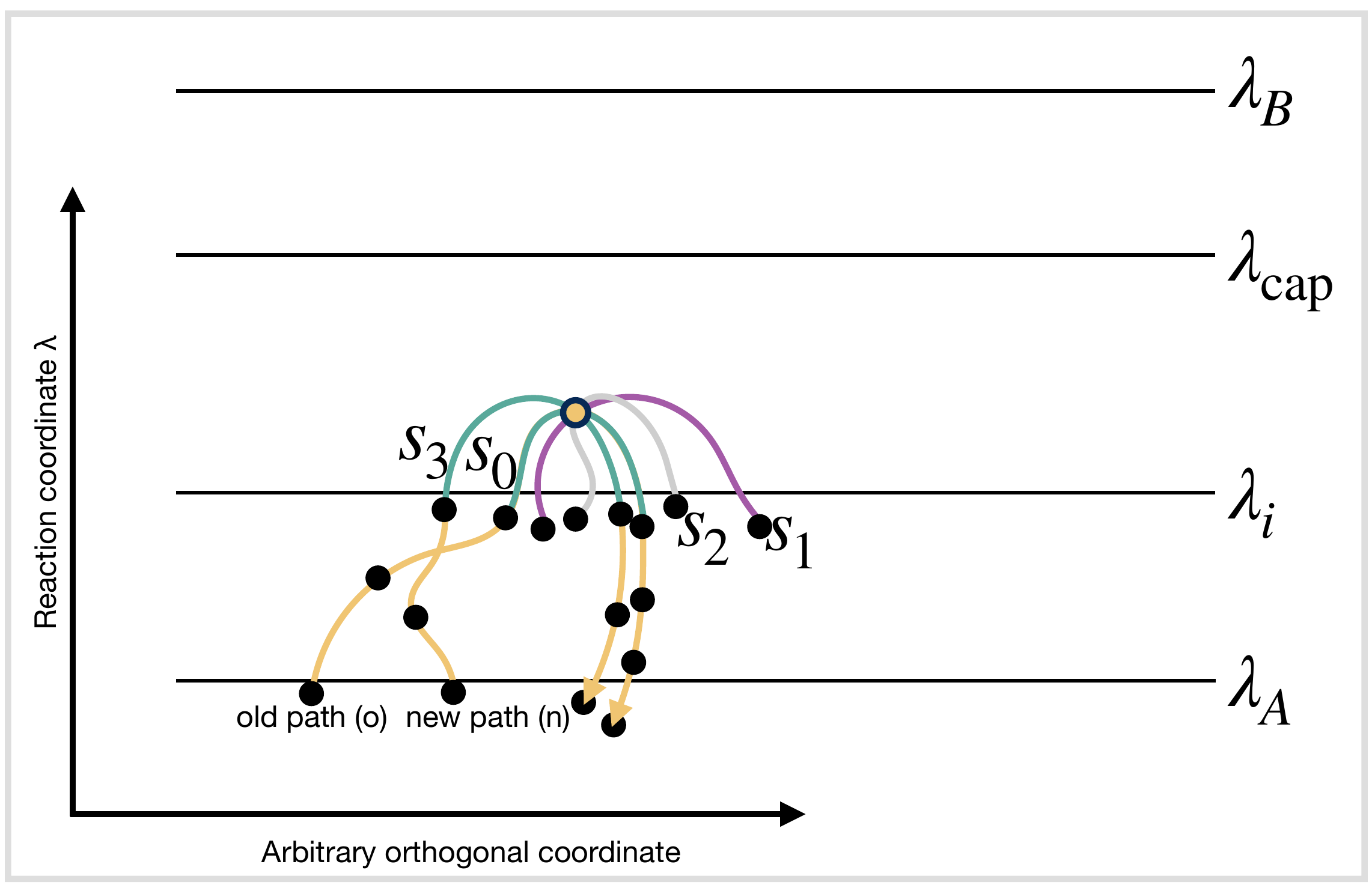}
\caption{
Illustration of a   wire-fencing  (WF) move in which multiple subtrajectories originate from the same shooting point.
This scarcity of shooting points can arise from infrequent frame saving, corresponding to a high subcycle number, while at the same time, steep gradients on the potential energy surface lead to a rapid decrease in the order parameter during shooting. Colors, markers, and line styles have the same meaning as in Fig.~\ref{fig:WF}.
}
\label{fig:oneshot}
\end{figure}

Although wire-fencing is generally more efficient than standard shooting, a high subcycle number combined with steep gradients on the potential energy surface,  can severely restrict the number of distinct shooting points available. In the worst-case scenario, multiple subtrajectories may therefore be initiated from the same shooting point, as illustrated in Fig.~\ref{fig:oneshot}.
This situation undermines one of the central advantages of WF: increasing the rate of decorrelation by avoiding shared phase points between consecutive full trajectories. 

A straightforward way to alleviate this issue would be to reduce the subcycle number. However, for some molecular systems, this may be difficult or computationally expensive, as it can substantially increase both storage requirements and input/output (I/O) overhead. For instance, the storage requirements associated with path-sampling simulations can become substantial for large biomolecular systems. A trajectory of a system containing $10^5$ atoms, saved every $1 \,\mathrm{ps}$, requires approximately $1 \, \mathrm{GB}$ per nanosecond, corresponding to roughly $1 \,\mathrm{TB}$ for a $1 \,\mu\mathrm{s}$ trajectory~\cite{Cheng2012}. Since path-sampling methods generate ensembles of trajectories rather than a single continuous trajectory, storing all sampled paths can consequently result in data volumes of several terabytes or more~\cite{Arjun2021}.

A related consideration arises in nucleation studies, where evaluating the order parameter can itself be computationally demanding because it often requires local structural descriptors to be calculated for a large fraction of the simulated system\cite{Dietrich2024}. Evaluating the order parameter at every MD step can therefore introduce substantial computational overhead. Consequently, it is often evaluated less frequently, provided that the sampling interval remains sufficiently short to resolve the relevant structural dynamics. For example, Haji-Akbari and Debenedetti evaluated a $q_6$-based order parameter every $1 \, \mathrm{ps}$ in their ice-nucleation simulations, corresponding to every 500 MD steps with a $2 \, \mathrm{fs}$ integration timestep~\cite{HajiAkbari2015}.

\subsection{General description of the MRWF move}
As discussed above, reducing the subcycle number increases storage requirements and computational cost. An alternative is to increase the frame-saving rate locally, for example around the $\lambda_i$ interface, while saving less frequently elsewhere.
However, this affects replica-exchange moves. Paths generated by shooting contain an increased density of shooting points around their respective interfaces.  After replica exchange, these densities are swapped between ensembles. Since the shooting points determine the distribution for subsequent shooting moves, the transition probability then depends not only on the current path but also on how it was generated.
This introduces a non-Markovian effect that can bias sampling even when detailed balance appears to be satisfied. Similar issues have recently been identified in other path-generation schemes.\cite{Falkner2025}
\begin{figure}[ht!]
\centering
\includegraphics[width=\linewidth]{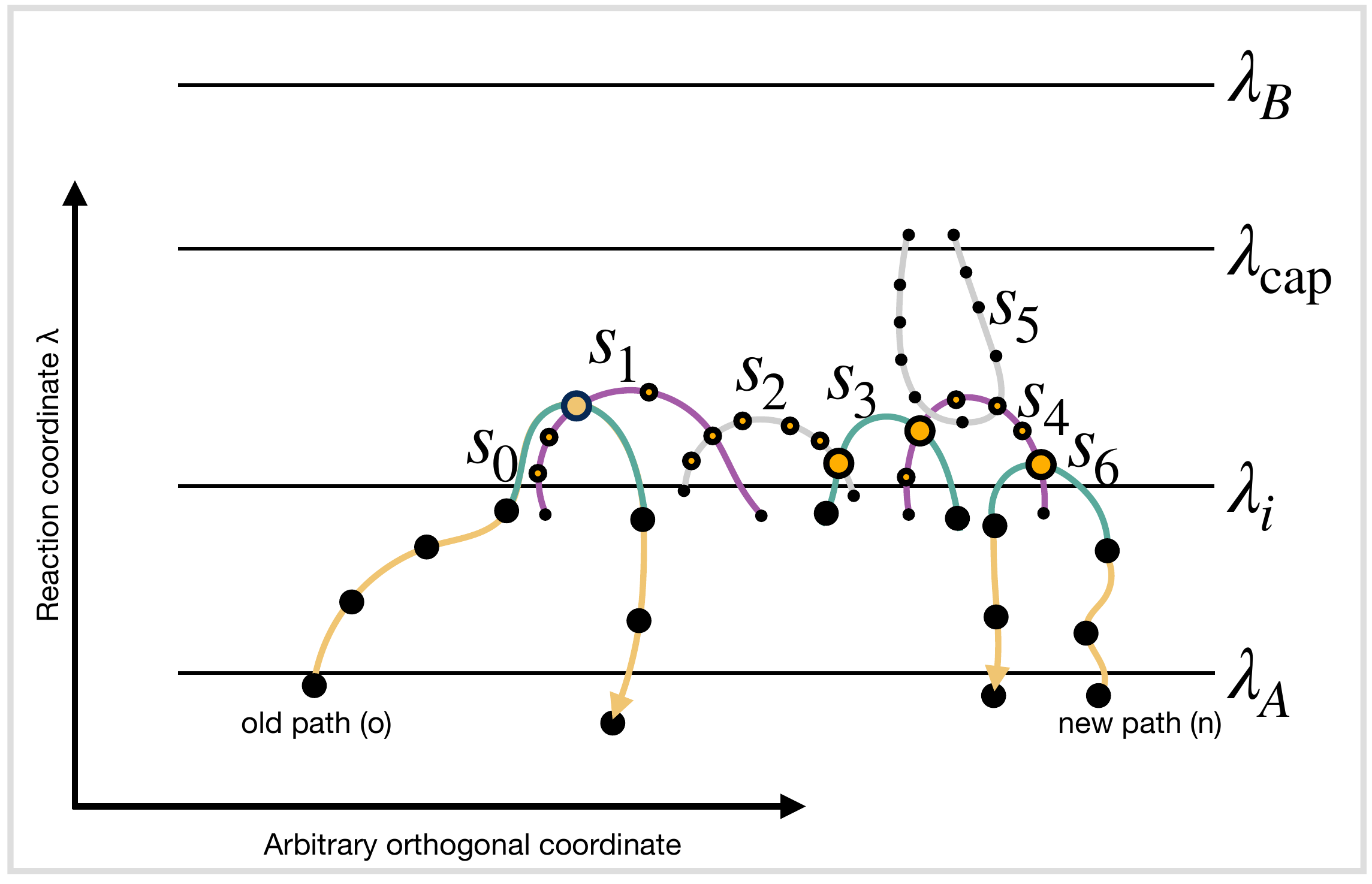}
\caption{
Illustration of the multi-resolution   wire-fencing  move. Colors, markers, and line styles have the same meaning as in Fig.~\ref{fig:WF}, with the smaller spheres representing the high-resolution points. This example illustrates how a new path is obtained from an old path using two sets of three subtrajectories. $s_5$ is a rejected subtrajectory, and $s_4$ is therefore used again to select a shooting point for generating $s_6$.
}
\label{fig:MRWF}
\end{figure}

The MC move proposed here exploits the additional freedom provided by subtrajectories in a  wire-fencing  move. Sampled paths are stored at low resolution, while selected, shorter subtrajectories are evaluated at higher resolution but not permanently stored. These form sets of subtrajectories, each terminating in a low-resolution subpath that can be used for extension to a complete path. The number of sets and subpaths per set are parameters of the MRWF algorithm.

Fig.~\ref{fig:MRWF} illustrates the MRWF move for two sets with three subpaths per set, 
showing how high-resolution subpaths circumvent the shooting-point scarcity discussed in the previous section.
 Although a single set with six subpaths may provide similar progression, multiple sets provide a safeguard against backtracking: if a final subpath is rejected, the algorithm can resume from the last accepted low-resolution subpath rather than discarding all the progress made by the preceding subtrajectories. Note that a high-resolution subpath cannot be used directly for extension, as this would violate the super-detailed-balance condition and lead to incorrect sampling.

For example, if $s_5$ is rejected because it terminates at $\lambda_{\rm cap}$ in both directions (as in Fig.~\ref{fig:MRWF}), the algorithm continues from $s_4$ by generating the low-resolution subpath $s_6$. If $s_6$ is also rejected, the extension instead proceeds from $s_3$, the last accepted low-resolution subpath. With only a single set, all generated subpaths would then be wasted and the extension would restart from $s_0$, producing a path strongly correlated with the original. In this case, rejecting the entire move would therefore be preferable. Multiple sets avoid this unnecessary loss of progress.

\subsection{The MRWF algorithm}
Below, we provide a step-by-step description of the MRWF algorithm. Here, $N_{\rm set}$ and $N_{\rm subpath}$ denote the number of sets and the number of subpaths per set, respectively, $N_{\rm succ}$ denotes the number of successfully completed sets, and $M(s_i)$ denotes the number of available shooting points for subpath $s_i$.
\vspace{-5ex}
\begin{center}
\captionof*{algorithm}{\bf }
\end{center}
\vspace{-2.em}
\begin{algorithmic}[1]
    \State By a weighted random selection, select segment $s_0$ from the old path $(o)$ that connects $\lambda_i$ to $\lambda_i$ or $\lambda_i$ to $\lambda_{\rm cap}$. This excludes segments of the form $\lambda_{\rm cap} \rightarrow \lambda_{\rm cap}$. The probability of selecting a given segment is proportional to the number of stored time slices within the $(\lambda_i, \lambda_{\rm cap})$ interval.
    \State Initialize $\text{countset}=1$,  $j=0$, $N_{\rm succ}=0$, and $\text{resolution}=\text{low}$.
    \State Select a shooting point from $s_j$ except endpoints.
    \State Generate new velocities from a Maxwell--Boltzmann distribution.
    \State Increment $j = j + 1$ 
    ($j$ is the index of the next subpath to be created).
    \State Set resolution:
    \Statex \hspace{\algorithmicindent} if $j < N_{\rm subpath}$: resolution = high
    \Statex \hspace{\algorithmicindent} if $j = N_{\rm subpath}$: resolution = low
    \State Generate $s_j$ by integrating backward and forward until $\lambda_i$ or $\lambda_{\rm cap}$.
    \State Check subpath type:
    \Statex \hspace{\algorithmicindent} if type $\lambda_{\rm cap}-\lambda_{\rm cap}$ or 
    \Statex 
\hspace{\algorithmicindent}
\hspace{\algorithmicindent}\hspace{-0.2em}
    ($M(s_j)$=1 and resolution=high):  
     go to step 9
    \Statex \hspace{\algorithmicindent} else go to step 10
    \State Reject subpath:
    \Statex \hspace{\algorithmicindent} if $j = 1$ or $j = N_{\rm subpath}$: go to step 11
    \Statex \hspace{\algorithmicindent} else set $s_j = s_{j-1}$ and repeat from step 3
    \State Accept subpath:
    \Statex \hspace{\algorithmicindent} if $j = N_{\rm subpath}$: 
    go to step 12
    \Statex \hspace{\algorithmicindent} else repeat from step 3
    \State Reject set: 
    \Statex \hspace{\algorithmicindent} let $s_j$ point to the last accepted low-resolution 
    \Statex \hspace{\algorithmicindent} path, $s_j = s_0$, and go to step 13
     \State Accept set:
    $N_{\rm succ}=N_{\rm succ}+1$
    \State Evaluate progress:
    \Statex \hspace{\algorithmicindent} 
    if $\text{countset} < N_{\rm set}$: 
    \Statex \hspace{\algorithmicindent} \hspace{\algorithmicindent} 
    $\text{countset}$=$\text{countset}+1$, set $s_0 = s_j$, $j = 0$
    \Statex \hspace{\algorithmicindent}
    \hspace{\algorithmicindent}
    return to step 3
     \Statex \hspace{\algorithmicindent}
     if $N_{\rm succ}=0$: go to step 17
    \State Extend subtrajectory: generate full trial trajectory using large $N_{\rm subcycle}$.
    \State If trial path is $B \rightarrow A$, reverse to get $A \rightarrow B$ path
    \State If trial path is valid: go to step 18
    \State Reject move:
    \Statex \hspace{\algorithmicindent} keep the old path: $(n) = (o)$ 
    \Statex \hspace{\algorithmicindent}
    go to step 19
     \State Accept move:
     $(n) =$  trial path
    \State Return path $(n)$.
\end{algorithmic}
Note that the index $j$ of subpath $s_j$ in the above algorithm is reset to zero after each set is completed. This differs from Fig.~\ref{fig:MRWF} and the discussion in the main text, where $j$ is treated as a continuously increasing counter to facilitate the discussion.

The above algorithm satisfies the super-detailed balance relation because, for each sequence of subtrajectories $\chi$, a corresponding reversible sequence $\overline{\chi}$ can be defined, analogous to the standard  wire-fencing  move. A complete move rejection occurs only when the trial path is found to be invalid at step 15. This can only occur for a $B \rightarrow B$ path and is expected to be rare, except for interfaces $\lambda_i$ located sufficiently high on the barrier.

At high resolution, temporarily storing every point of a subtrajectory is unnecessary, since only a limited number will be used for subsequent shooting attempts. The required number equals the number of remaining shooting attempts in a set. Therefore, in addition to the store-all method, we implemented a memory-efficient reservoir-sampling variant of the algorithm that stores only these points. For $n$ required shooting points, the algorithm initially stores $n$ copies of the first point. 
As the subtrajectory grows, the newly generated time slice replaces each stored point with probability $1/m$, where $m$ is the current length of the subtrajectory in time slices.
Consequently, at the end of the subtrajectory, the $n$ stored points constitute a uniform sample of the entire subtrajectory, yielding shooting-point selection statistics exactly equivalent to those of the store-all approach.

\subsection{MRWF and super-detailed balance}

We refer the reader to Ref.~\citenum{WF} for a detailed derivation of super-detailed balance for subtrajectory moves and   wire-fencing  in particular. The same principles apply here, with additional considerations arising from the use of both low- and high-resolution subtrajectories. These are discussed below, together with the steps of the algorithm that require further justification.

The rejection at step 8 when $M(s_j)=1$ and the resolution is high is not required for super-detailed balance; the subpath could equally well be accepted. However, such a subpath provides only a single shooting point despite a low subcycle, thereby reintroducing the shooting-point scarcity discussed above. Rejecting the subpath instead allows the next shooting point to be selected from the preceding subpath, which may provide a more suitable point farther from the $\lambda_i$ interface. 

Thus, rejection at step 8 is an efficiency-based algorithmic choice, as both acceptance and rejection satisfy super-detailed balance. This differs from rejecting the set at step 9 and proceeding to step 11 when $j=1$ or $j=N_{\rm subpath}$. In standard wire fencing, the next subpath would simply be generated from the previously generated one. In MRWF, however, this would violate super-detailed balance because the change in resolution would break the required reversibility between $\chi$ and $\overline{\chi}$.

To illustrate this point, consider a single set, $N_{\rm set}=1$, containing $N_{\rm subpath}=3$ subpaths. Suppose that, after the subtrajectory $s_1$ is rejected, the algorithm nevertheless continues by generating the next subtrajectory $s_2$ from $s_0$, following the standard  wire-fencing  procedure. If the subsequent subtrajectories are accepted, the resulting sequence $\chi$ can be represented as
\[
\begin{aligned}
\chi &=
\begin{tikzcd}[column sep=large, baseline=14pt]
s_0 \arrow[r] \arrow[d, shift right=2] & s_2 \arrow[r] & s_3 \\
s_1 \arrow[u, shift right=2]
\end{tikzcd}
\end{aligned}
\]
and the corresponding reverse move $\overline{\chi}$ would be
\[
\begin{aligned}
\overline{\chi} &=
\begin{tikzcd}[column sep=large, baseline=14pt]
s_3 \arrow[r] & s_2 \arrow[r] & s_0 
\arrow[d, shift right=2]\\
            & & s_1 \arrow[u, shift right=2]
\end{tikzcd}
\end{aligned}
\]

This construction is reversible for standard wire fencing but not for MRWF, where the first and last subpaths must be generated at low resolution and the intermediate subpaths at high resolution. Thus, for $\chi=s_0,s_1,s_2,s_3$, $s_0$ and $s_3$ are low resolution, while $s_1$ and $s_2$ are high resolution. The reverse sequence, $\overline{\chi}=s_3,s_2,s_0,s_1$, does not preserve this resolution ordering and is therefore not reversible. Consequently, if the first subpath in a set is rejected, the entire set must be rejected rather than continuing from the last accepted subpath.

Now, let us further examine the purpose of sets. Suppose that, in Fig.~\ref{fig:MRWF}, the 6th subpath is rejected because it terminates at $\lambda_{\rm cap}$ in both directions. The rejected subpath is not used for extension, as this would likely produce a $B\rightarrow B$ path. Unlike standard wire fencing, however, the construction cannot simply continue from the last accepted subpath. Since this subpath has high resolution, extending it would violate super-detailed balance. Instead, the trajectory is extended from the last accepted low-resolution subpath, preserving the required resolution ordering and reversibility between the forward and reverse constructions.

To illustrate this, consider an MRWF move with the following parameters:
$N_{\rm set}=2$ and $N_{\rm subpath}=3$.
Suppose that all subtrajectories are successful except for the last one.
The resulting sequence $\chi$ is then
\[
\begin{aligned}
\chi &=
\begin{tikzpicture}[
    >=stealth,
    node distance=1.cm,
    baseline=(s0.base)
]
    \node (s0) {$s_0$};
    \node (s1) [right of=s0] {$s_1$};
    \node (s2) [right of=s1] {$s_2$};
    \node (s3) [right of=s2] {$s_3$};
    \node (s4) [below of=s3] {$s_4$};
    \node (s5) [below of=s4] {$s_5$};
    \node (s6) [below of=s5] {$s_6$};

    \draw[->] (s0) -- (s1);
    \draw[->] (s1) -- (s2);
    \draw[->] (s2) -- (s3);
    \draw[->] (s3) -- (s4);
    \draw[->] (s4) -- (s5);
    \draw[->] (s5) -- (s6);

    \draw[->] (s6) to[out=60,in=300,looseness=1.5] (s3);
\end{tikzpicture}
\end{aligned}
\]
In this case, $s_3$, which is stored at low resolution, is eventually extended to generate a new path, so only part of the computational effort is lost while the progression $s_0\to s_1\to s_2\to s_3$ still contributes to decorrelating the new path from the old one. In contrast, with $N_{\rm set}=1$ and $N_{\rm subpath}=6$, rejection of the final subpath discards the entire computational effort of the move. Multiple sets therefore safeguard against unproductive MD integration, particularly for interfaces high up the barrier and large numbers of subpaths per move, $N_{\rm set}\times N_{\rm subpath}$.

The above sequence $\chi$ is reversible, as required to satisfy the
super-detailed balance relation:
\[
\begin{aligned}
\overline{\chi} &=
\begin{tikzpicture}[
    >=stealth,
    node distance=1.cm,
    baseline=(s0.base)
]
    \node (s0) {$s_3$};
    \node (s1) [right of=s0] {$s_2$};
    \node (s2) [right of=s1] {$s_1$};
    \node (s3) [right of=s2] {$s_0$};
    \node (s4) [below of=s0] {$s_4$};
    \node (s5) [below of=s4] {$s_5$};
    \node (s6) [below of=s5] {$s_6$};

    \draw[->] (s0) -- (s1);
    \draw[->] (s1) -- (s2);
    \draw[->] (s2) -- (s3);
    \draw[->] (s0) -- (s4);
    \draw[->] (s4) -- (s5);
    \draw[->] (s5) -- (s6);

    \draw[->] (s6) to[out=60,in=300,looseness=1.5] (s0);

\end{tikzpicture}
\end{aligned}
\]

The reverse sequence $\overline{\chi}=s_3, s_4, s_5, s_6; s_2, s_1, s_0$ represents a move initiated from the existing path segment $s_3$. The third subtrajectory, $s_6$, is rejected, causing the first set to be rejected. The second set is then generated without rejection, ending with the low-resolution subpath $s_0$, which is subsequently extended to form the full path.

It might seem attractive to apply an algorithmic choice similar to step 8 and reject the entire move when the first set is rejected, since the procedure would in either case have to restart from a segment of the original path, whether by continuing with the second set or by rejecting and initiating a new MRWF move. The latter would allow a different segment to be selected (the old path $(o)$ in Fig.~\ref{fig:WF}, for example, contains three possible segments). The sequences $\chi$ and $\overline{\chi}$ above show that this choice would break reversibility unless the entire move were also rejected whenever the last set is rejected. This would defeat the purpose of using multiple sets, namely, to avoid discarding computationally expensive MD steps following a rejection.
 On the other hand, at step 13, the full MC move is rejected if none of the sets is successful.  Rejecting the move avoids extending the initial subtrajectory $s_0$  extracted from the old path, which would likely produce a new path that is strongly correlated with the old path. 

In summary, we have shown that the concept of super-detailed balance underlying the  wire-fencing  move can be generalized to the more complex MRWF move, provided that the additional requirements for reversibility arising from the use of two resolutions are properly accounted for. In MRWF, rejection of a subpath does not necessarily result in continuation from the immediately preceding subpath, as the move may instead backtrack across several subtrajectories. This can lead to substantial computational waste. Introducing multiple sets mitigates this issue by allowing the procedure to resume from a lower-resolution subpath, thereby preserving reversibility while retaining useful progress from most of the preceding subtrajectories within a single move. MRWF can therefore be an effective alternative to wire fencing, particularly when storage requirements or the computational overhead of order-parameter evaluation constitute significant bottlenecks.

\section{Results}

We implemented MRWF in the $\infty$RETIS code, \texttt{infRETIS} \cite{infretissoftware}, and directly compared its accuracy and performance with those of conventional wire-fencing across three systems of increasing complexity.
The production $\infty$RETIS simulations followed the protocol described by Zhang \textit{et al.}~\cite{PNAS2024}, with conventional shooting applied in the $[0^-]$ and $[0^+]$ ensembles and high-acceptance shooting moves applied in the remaining positive ensembles, using either wire-fencing or MRWF. 

For each of the three systems, MRWF and wire-fencing simulations were performed using otherwise identical 
settings. In particular, the same molecular model, ensemble definitions, interfaces, and low-resolution subcycle settings were used for the corresponding MRWF and wire-fencing calculations. MRWF additionally used a high-resolution subcycle during the intermediate subtrajectory construction. The low-resolution MRWF subcycle was set equal to the subcycle used in the corresponding wire-fencing simulation, such that the temporal resolution of the final stored trajectories was identical for the two methods.

To provide a comparable subtrajectory-generation budget, the number of subtrajectories per move was fixed at six for both methods. For MRWF, these were organized into two sets of three subtrajectories each ($N_{\rm set}=2$ and $N_{\rm subpath}=3$), whereas conventional wire-fencing used six successive subtrajectories.

Rate constants were obtained from the standard TIS/RETIS relation $k_{AB}=f_AP_A(\lambda_B|\lambda_A)$~\cite{TIS,PNAS2024}, where $f_A$ is the effective positive flux and $P_A(\lambda_B|\lambda_A)$ is the probability of reaching state $B$ after crossing $\lambda_A$ before returning to state $A$. 
The crossing probabilities were obtained from the sampled path ensembles using a variant\cite{REW2026} of path reweighting~\cite{Rogal2010,vanErp2016,REW2026}, which accounts for the high-acceptance weights and fractional sampling counts that arise from the infinite-swapping algorithm.
Statistical uncertainties were estimated using recursive block error analysis~\cite{PyRETIS3}. The reported uncertainties correspond to one standard error of the mean (SEM), obtained by averaging the SEM estimates over a plateau region of the block-length range, where the estimated uncertainty is approximately independent of the block length.

\subsection{One-dimensional Langevin particle in a double-well}

To assess whether the MRWF move reproduces established results for a simple benchmark\cite{vanErp2012} system with highly converged statistics, we studied the one-dimensional double-well system using \texttt{infRETIS}'s internal MD engine~\cite{TurtleMD}, which employs a Langevin-inertia integrator. The order parameter $\lambda$ was given by the particle position along the $x$ coordinate. State $A$ was defined at $\lambda_A=-0.99$ and state $B$ at $\lambda_B=1.0$. 
The MD integration time step was $\Delta t = 0.025$ in reduced units, and the temperature was set to $T=0.07$. The Boltzmann constant and particle mass were set to unity, $k_{\mathrm B}=m=1$, and the Langevin friction coefficient was $\gamma=0.3$.

For both wire-fencing and MRWF, the low-resolution subcycle was set to 5. In MRWF, the high-resolution subtrajectories were propagated using a subcycle of 1. The path sampling simulations were run for a total of $1,000,000$ 
Monte Carlo moves, which represent the number of attempted path-generation moves, including point-exchange, standard shooting, wire-fencing, or MRWF.

Fig.~\ref{fig:main_results}(a,b) shows crossing-probability profiles and running rate estimates for both MRWF and WF, which yield essentially identical results. The endpoint inset in
Fig.~\ref{fig:main_results}(a) resolves the small numerical difference
between the otherwise overlapping crossing-probability profiles. The
final rates were $(2.49\pm0.06)\times10^{-7}$ 
for
MRWF and $(2.48\pm0.06)\times10^{-7}$ 
for WF,
while the corresponding crossing probabilities were
$(5.65\pm0.13)\times10^{-7}$ and $(5.64\pm0.14)\times10^{-7}$,
respectively (Table~\ref{tab:results_summary}). 
\begin{table*}[ht!]
    \centering
    \small
    \caption{Summary of final path-reweighting estimates. All uncertainties are one SEM from block error analysis. Units are given directly with the rate and flux estimates; r.u. denotes reduced units. The crossing probability $P_{A}(\lambda_B|\lambda_A)$ is dimensionless.}
    \label{tab:results_summary}
    \setlength{\tabcolsep}{3pt}
    \begin{tabular*}{\textwidth}{@{\extracolsep{\fill}}llccc@{}}
        \toprule
        System & Method & $k$ & $f_A$ & $P_{A}(\lambda_B|\lambda_A)$ \\
        \midrule
        Double well & wire-fencing   & $2.48 \times 10^{-7} \pm 6.08 \times 10^{-9}\,\mathrm{r.u.}$ & $4.40 \times 10^{-1} \pm 4.22 \times 10^{-4}\,\mathrm{r.u.}$ & $5.64 \times 10^{-7} \pm 1.39 \times 10^{-8}$ \\
        Double well & MRWF & $2.49 \times 10^{-7} \pm 5.65 \times 10^{-9}\,\mathrm{r.u.}$ & $4.40 \times 10^{-1} \pm 3.39 \times 10^{-4}\,\mathrm{r.u.}$ & $5.65 \times 10^{-7} \pm 1.28 \times 10^{-8}$ \\
        Puckering   & wire-fencing   & $3.53 \times 10^{5} \pm 2.24 \times 10^{4}\,\mathrm{s}^{-1}$ & $1.20 \times 10^{13} \pm 1.93 \times 10^{11}\,\mathrm{s}^{-1}$ & $2.94 \times 10^{-8} \pm 1.80 \times 10^{-9}$ \\
        Puckering   & MRWF & $3.44 \times 10^{5} \pm 2.61 \times 10^{4}\,\mathrm{s}^{-1}$ & $1.18 \times 10^{13} \pm 1.62 \times 10^{11}\,\mathrm{s}^{-1}$ & $2.91 \times 10^{-8} \pm 2.15 \times 10^{-9}$ \\
        T4L L99A / benzene & wire-fencing   & $24.7 \pm 39.8\,\mathrm{s}^{-1}$ & $1.92 \times 10^{10} \pm 3.97 \times 10^{9}\,\mathrm{s}^{-1}$ & $1.28 \times 10^{-9} \pm 3.38 \times 10^{-9}$ \\
         T4L L99A / benzene & MRWF & $227 \pm 97\,\mathrm{s}^{-1}$ & $1.82 \times 10^{10} \pm 3.60 \times 10^{9}\,\mathrm{s}^{-1}$ & $1.25 \times 10^{-8} \pm 6.00 \times 10^{-9}$ \\
        \bottomrule
    \end{tabular*}
\end{table*}
These values are also
consistent with previous results for the same double-well benchmark,
including the earlier\cite{vanErp2012} RETIS rate estimate of
$(2.79\pm0.70)\times10^{-7}$, the $\infty$RETIS/WF result~\cite{InfRET1} of
$(2.59\pm0.07)\times10^{-7}$, and the approximate Kramers-theory value
of $2.58\times10^{-7}$~\cite{vanErp2012,InfRET1}.

\subsection{Atomistic Puckering System}

To assess the validity of the MRWF move in a more complex atomistic system with explicit solvent, we applied both  wire-fencing  and MRWF to a puckering system consisting of one oxane molecule solvated by 291 TIP3P water molecules. This system is substantially more complex than the one-dimensional system considered in the previous subsection, while still allowing sufficient convergence to draw reliable conclusions about the accuracy of MRWF relative to  wire-fencing. Oxane was described using the OpenFF 2.1 force field~\cite{Boothroyd2023Sage,OpenFF210}, and simulations were performed with GROMACS~\cite{Abraham2015GROMACS}.

The puckering coordinates were calculated using the Cremer--Pople representation for a six-membered ring, described by the coordinates $\theta$, $\phi$, and the puckering amplitude $Q$. The scalar order parameter used for path sampling was the polar angle $\theta$. State $A$ was defined at $\theta=10^\circ$, the cap interface was placed at $\lambda_{\mathrm{cap}}=70^\circ$, and state $B$ at $\theta=90^\circ$.

Simulations were performed at $300\,\mathrm{K}$ using an MD integration time step of $2\,\mathrm{fs}$. For both wire-fencing and MRWF, the low-resolution subcycle was set to 4, while the MRWF high-resolution intermediate subtrajectories were propagated using a subcycle of 1. The simulations were run for a total of $250,000$ Monte Carlo moves.

As in the previous one-dimensional Langevin system, 
Fig.~\ref{fig:main_results}(c,d) shows that MRWF and wire-fencing yield nearly overlapping
crossing-probability profiles and converge to nearly the same rate constant in this
more complex atomistic system. Although the endpoint inset in
Fig.~\ref{fig:main_results}(c) reveals a small difference between the final
crossing probabilities, the resulting rate estimates remain effectively
identical.
The final rates were
$(3.44\pm0.26)\times10^{5}\,\mathrm{s}^{-1}$ for MRWF and
$(3.53\pm0.22)\times10^{5}\,\mathrm{s}^{-1}$ for WF, with crossing
probabilities of $(2.91\pm0.22)\times10^{-8}$ and
$(2.94\pm0.18)\times10^{-8}$, respectively
(Table~\ref{tab:results_summary}). Thus, the use of higher-resolution
intermediate subtrajectories in MRWF leaves the sampled kinetic
observables unchanged within statistical uncertainty also for an
explicit-solvent atomistic system.

\subsection{T4L L99A/benzene ligand unbinding}

Having established agreement in the two validation systems, we next consider a protein--ligand unbinding process in the T4L L99A/benzene complex, where the large subcycle probes the regime for which MRWF was designed. The T4 lysozyme L99A/benzene complex was based on the crystallographic structure with PDB code 1L84~\cite{Eriksson1992}. The simulation files were obtained from PLUMED-NEST~\cite{PLUMEDConsortium2019}, plumID:23.030 ~\cite{Ray2024}.
System preparation followed the T4L/benzene setup described by Capelli \textit{et al}~\cite{Capelli2019}. The protein was described using the AMBER14SB force field, the solvent using the TIP3P water model, and benzene using the General AMBER Force Field (GAFF). 

The system was first energy-minimized for $2,000$ steps using the steepest-descent algorithm.
Following minimization, the system was equilibrated for $2\,\mathrm{ns}$ in the NVT ensemble, followed by $5\,\mathrm{ns}$ in the NPT ensemble at a pressure of $1\,\mathrm{bar}$. A further $50\,\mathrm{ns}$ MD equilibration simulation was subsequently performed, using a timestep of 2 fs in all cases.

The order parameter $\lambda$ was defined as the average minimum distance between the ligand and five protein residues: Ile78, Tyr88, Met102, Val111, and Leu121. These residues were selected as those having, on average, the smallest minimum distance to the ligand during the preceding equilibrium MD simulation. State $B$ was defined at an average minimum distance of $2\,\mathrm{nm}$ between the ligand and these residues.

\begin{figure*}[t]
    \centering
    \includegraphics[width=0.95\textwidth]{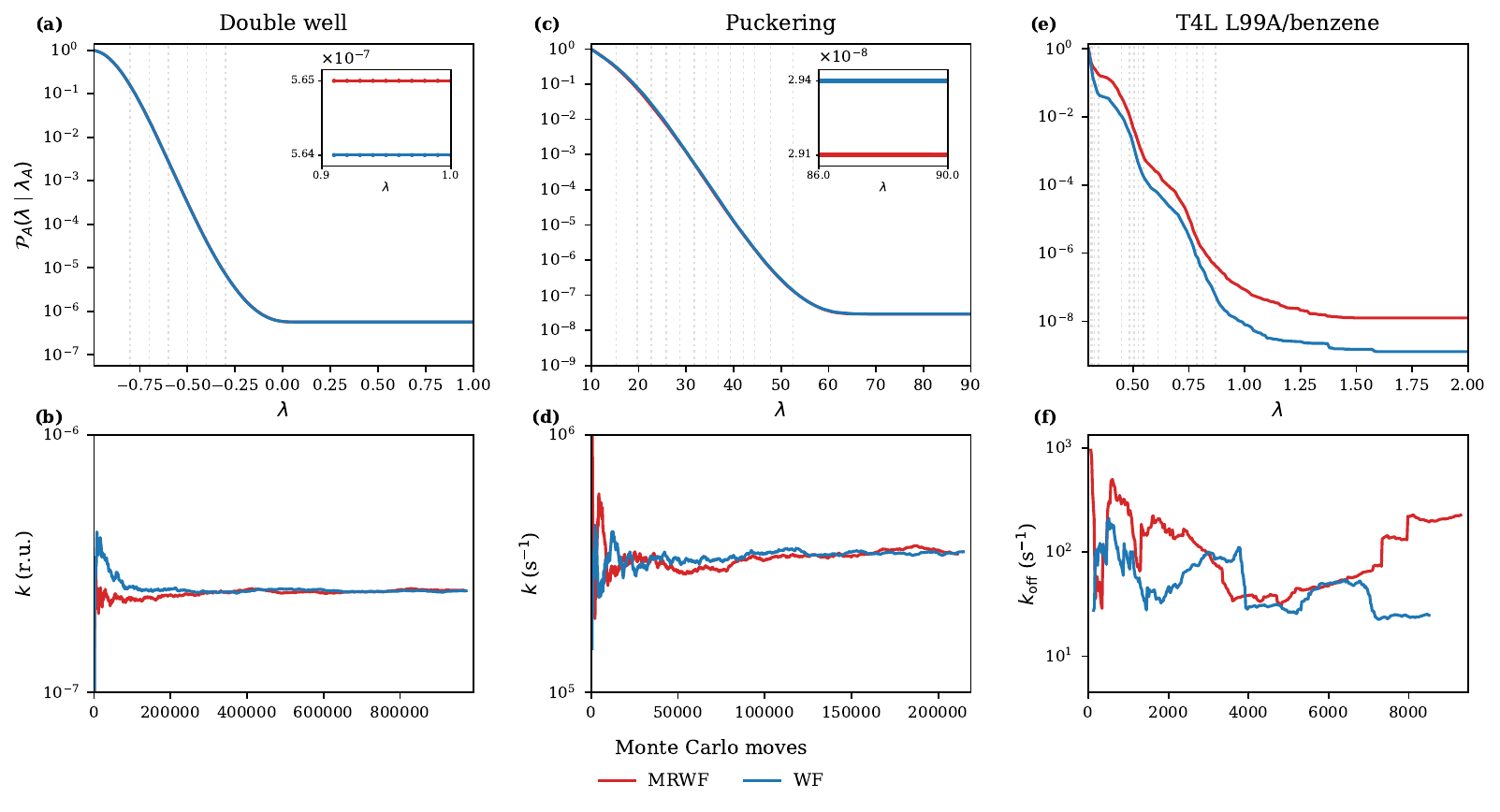}
\caption{Comparison of MRWF and conventional  wire-fencing  for the double-well, puckering, and T4L L99A/benzene systems. Panels (a,b), (c,d), and (e,f) correspond to the double-well, puckering, and T4L systems, respectively; within each pair, the first panel shows the crossing-probability profile and the second the corresponding running rate estimate as a function of Monte Carlo moves. Insets in (a) and (c) show the endpoint region of the double-well and puckering crossing-probability profiles on a linear scale. The vertical dashed lines in panels (a), (c), and (e) represent the interface positions.}
    \label{fig:main_results}
\end{figure*}

$\lambda_A=0.3\,\mathrm{nm}$ and $\lambda_B=2.0\,\mathrm{nm}$ were specified beforehand. The intermediate interfaces were placed iteratively with the \texttt{inf-init} initialization algorithm~\cite{infretissoftware}, such that the conditional crossing probabilities,
$
P_A(\lambda_{i+1}\mid\lambda_i),
$
were approximately $0.2$--$0.3$ between neighboring interfaces. The resulting setup consisted of 16 ensembles, with 14 intermediate interfaces placed by \texttt{inf-init}. The final interface positions are shown as dashed vertical lines in Fig.~\ref{fig:main_results}(e).

Prior to the production simulations, wire-fencing and MRWF were initialized independently and run for 14 days using four workers sharing one NVIDIA A100 GPU. After this initialization period, the two runs were compared based on their preliminary crossing-probability profiles and resulting interface placement. The MRWF initialization produced more reactive paths and a better-resolved crossing-probability profile than the corresponding wire-fencing initialization, and its interface set and sampled paths were therefore selected as the common starting point for both production simulations. Using the same interface set and initial paths ensured that differences observed between wire-fencing and MRWF could be attributed to the production sampling rather than to different interface placements or initial path configurations.

The production $\infty$RETIS simulations were then run independently for 20 days using 8 workers for both WF and MRWF respectively. Each simulation used two NVIDIA L40S GPUs, with 16 CPU cores available per GPU. For T4L L99A/benzene, one stored low-resolution frame corresponded to $1000\times0.002\,\mathrm{ps}=2\,\mathrm{ps}$, and rates and fluxes were therefore reported in $\mathrm{s}^{-1}$.

As shown in Fig.~\ref{fig:main_results}(e,f) and Table~\ref{tab:results_summary}, convergence is substantially more challenging for both MRWF and wire-fencing than for the two previously studied systems. The two methods yield comparable flux estimates, $1.82\times10^{10}\,\mathrm{s}^{-1}$ and $1.92\times10^{10}\,\mathrm{s}^{-1}$, respectively. However, their crossing-probability profiles progressively diverge with increasing interface index (Fig.~\ref{fig:main_results}(e)). The final crossing probabilities were $(1.25\pm0.60)\times10^{-8}$ for MRWF and $(1.28\pm3.38)\times10^{-9}$ for WF, differing by approximately one order of magnitude (Table~\ref{tab:results_summary}). This separation is reflected directly in the running rate estimates (Fig.~\ref{fig:main_results}(f)), which reached $227\pm97\,\mathrm{s}^{-1}$ for MRWF and $24.7\pm39.8\,\mathrm{s}^{-1}$ for WF.

The separation between the wire-fencing and MRWF crossing-probability profiles becomes pronounced in the region spanned by ensembles $[5^+]$--$[7^+]$, corresponding to interface positions $\lambda=0.449$--$0.504\,\mathrm{nm}$ (Fig.~\ref{fig:main_results}(e)). Across this region, the wire-fencing profile continues to decrease more steeply, whereas MRWF develops a smoother tail and retains a higher probability of progressing toward the product state. As shown below, this is also the region in which repeated shooting-point selection becomes prevalent for WF. 

For reference, the corresponding residence times were obtained as $\tau_{\mathrm{res}}=1/k_{\mathrm{off}}$, giving approximately $4.4,\mathrm{ms}$ for MRWF and $40.5,\mathrm{ms}$ for WF. A previous atomistic study using the same molecular model reported a dissociation timescale of approximately 2.5 ms~\cite{Ray2024}, while the experimental dissociation rate of $950~\mathrm{s}^{-1}$ corresponds to a residence time of approximately 1.05 ms~\cite{Feher1996}. Given the statistical uncertainty of the simulations, uncertainties in the experimental estimate, and possible systematic effects associated with the force field, model, and state definitions, this comparison should be regarded as an external consistency check rather than as a quantitative reproduction of the experimental rate. Nevertheless, the closer agreement of MRWF with the previous atomistic estimates provides additional support for the MRWF result relative to the substantially longer residence time obtained with wire-fencing.

More generally, underestimation of rare-event rates is a common concern in rare-event sampling when exploration of degrees of freedom orthogonal to the chosen order parameter is insufficient. In such cases, a more favorable reaction pathway may remain unexplored, leading to an overestimated residence time. Overestimation of the rate, corresponding to an underestimated residence time, is less readily produced by incomplete exploration because it would require preferential sampling of an unusually fast pathway. It can nevertheless arise, for example, from inadequate initialization or other systematic sampling effects. Such an initialization bias would typically be detectable in the running rate estimate shown in Fig.~\ref{fig:main_results}(f), which would be expected to start at an anomalously high value and subsequently decrease as the initially biased trajectories are replaced by properly sampled ones. In the present simulations, we used \texttt{inf-init} to generate high-quality initial paths, and none of the path-generation moves performed during the initialization stage were included in the production data. We therefore consider an initialization-induced overestimation of the rate unlikely in this case.

In the present case, the smoother MRWF curve of the crossing probability compared with the wire-fencing result in Fig.~\ref{fig:main_results}(e) provides further qualitative evidence that MRWF achieves more consistent sampling. Taken together with the agreement with previous atomistic estimates, these observations increase our confidence that the MRWF result provides a more reliable estimate of the dissociation rate for this system.

\subsection{Computational efficiency}

For each system, the corresponding wire-fencing and MRWF simulations used the same computational resources. As a measure of raw sampling throughput, we report the number of sampled paths generated per unit wall-clock time. However, this quantity alone does not necessarily indicate which method is more statistically efficient, since a large number of highly correlated paths may provide less statistical information than a smaller number of highly decorrelated paths. We therefore additionally quantify the statistical efficiency of the rate estimate as $\eta = 1/(\epsilon^2 T)$, where $T$ is the wall-clock time and $\epsilon$ is the relative statistical error of the rate estimate. The statistical errors were obtained using recursive block error analysis, such that this measure accounts for correlations between sampled paths to the extent that these correlations are reflected in the estimated uncertainty.

For the double-well system, MRWF required approximately $38\%$ more wall-clock time than wire-fencing to generate the same number of sampled paths. This additional computational cost was partly offset by the lower statistical error obtained with MRWF, with relative errors in the final rate estimates of approximately $2.27\%$ for MRWF and $2.45\%$ for WF. Nevertheless, the resulting statistical efficiency, $\eta$, was approximately $15.5\%$ lower for MRWF than for  wire-fencing, as estimated from $\eta_{\mathrm{MRWF}}/\eta_{\mathrm{WF}} = (2.45)^2/[1.38(2.27)^2] \approx 0.845$.

The trend in wall-clock time and relative statistical error between MRWF and  wire-fencing  is reversed for the puckering system. MRWF completes the simulation in 8.3 days, compared with 8.7 days for WF, corresponding to approximately $5\%$ less wall-clock time for generating the same number of sampled paths. However, the corresponding relative errors in the rate estimates are approximately $7.59\%$ and $6.35\%$ for MRWF and WF, respectively. Consequently, the statistical efficiency is lower for MRWF, with $\eta_k^{\mathrm{MRWF}}/\eta_k^{\mathrm{WF}}\approx0.73$, corresponding to a $27\%$ reduction relative to WF. Although these differences in efficiency are modest, they indicate that MRWF does not improve the statistical efficiency of the rate estimate relative to wire-fencing for the simple systems considered here, which do not require a high subcycle number.

This is in striking contrast to the T4L L99A/benzene ligand-unbinding study, where the combination of a large subcycle number and steep regions in the potential energy surface leads to pronounced differences in the generation of unique shooting points. The two production runs have comparable wall-clock times of approximately 21 days, but Table~\ref{tab:results_summary} shows that the relative errors in the final rate estimates are approximately $43\%$ for MRWF and $161\%$ for WF. With approximately equal wall-clock times, this corresponds to a rate-estimation efficiency $\eta$ that is approximately 14-fold higher for MRWF than for WF. When the analysis is instead based on the crossing probability, the difference is even more pronounced. The corresponding relative errors in the crossing-probability estimates are approximately $48\%$ for MRWF and $264\%$ for WF, yielding an approximately 30-fold higher statistical efficiency for MRWF.
These results should, however, be interpreted with care because of the uncertainties in these error estimates.

The raw sampling throughput shows a smaller difference, with MRWF generating $11.7\%$ more sampled paths over comparable wall-clock times: 9180 accepted paths for MRWF compared with 8218 for WF. We attribute this difference in throughput to the earlier termination of MRWF trajectories when the terminating interfaces, $\lambda_i$ and $\lambda_{\rm cap}$, are crossed at high resolution. In contrast, the  wire-fencing  subpaths have a minimum length of 2000 MD steps, since both the forward and backward segments contain at least 1000 MD steps, corresponding to the subcycle length. 

Thus, the high-resolution subpaths gives MRWF a modest advantage in computational throughput. However, this effect is not the primary reason for its substantially higher statistical efficiency.
The substantially larger gain in statistical efficiency therefore indicates that the improvement is not primarily due to generating more paths, but rather to obtaining a lower statistical error per unit wall-clock time, consistent with greater decorrelation between sampled paths.

\subsection{Shooting-point analysis}
To investigate whether the difference between wire-fencing and MRWF is related to differences in shooting-point selection, we examined how often the same shooting point was selected in successive subpath generations. Although this metric is calculated per wire-fencing or MRWF move, a high fraction of repeated shooting points also increases the likelihood that two or more successive full trajectories share a common phase point. Such an observation provides a direct visual indication that the corresponding full trajectories are strongly correlated. However, successive full trajectories may still exhibit substantial correlation even when they do not share a common phase point, particularly when the shooting point changes only a few times during an advanced shooting move. Thus, while the frequency of repeated shooting-point selections characterizes the path-generation move itself, it also provides an indication of the degree of decorrelation achieved between full sampled trajectories over successive advanced shooting realizations.

\begin{table}[hb]
\centering
\small
\caption{Repetition of successive shooting-point selections in accepted wire-fencing and MRWF moves. A shooting point is counted as repeated only when both its selected index and order-parameter value are unchanged from the preceding selection. For MRWF, all successive low- and high-resolution shooting-point selections are included.}
\label{tab1}
\begin{tabular}{lcc}
\hline
Ensemble(s) & WF repeats & MRWF repeats \tabularnewline
\hline
$[1^+]$--$[15^+]$ & 25.2\% & 1.6\% \tabularnewline
$[1^+]$--$[4^+]$ & 0.6\% & 0.1\% \tabularnewline
$[5^+]$ & 50.0\% & 3.2\% \tabularnewline
$[6^+]$ & 53.5\% & 4.1\% \tabularnewline
$[7^+]$ & 42.5\% & 3.5\% \tabularnewline
$[8^+]$--$[15^+]$ & 28.0\% & 1.6\% \tabularnewline
\hline
\end{tabular}
\end{table}
As expected, Table~\ref{tab1} shows that repeated selection of identical shooting points occurs substantially more often in wire-fencing than in MRWF. Averaged over all $[i^+]$ ensembles with $i>0$, repeated selections occur in 25.2\% of wire-fencing subpath generations, compared with only 1.6\% for MRWF. The $[0^-]$ and $[0^+]$ ensembles are sampled using standard shooting moves and are therefore not included in this analysis.

The ensemble-resolved statistics provide a more detailed picture. As shown in Table~\ref{tab1}, the first ensembles, up to $[4^+]$, exhibit very few repeated shooting-point selections for either method. In contrast, for wire-fencing, repeated selections occur in approximately half of the subpath generations in the $[5^+]$, $[6^+]$, and $[7^+]$ ensembles. For MRWF, the corresponding fraction remains below 5\% for all ensembles. In the subsequent $[8^+]$--$[15^+]$ ensembles, the repetition rate for wire-fencing decreases again, but remains substantial, with approximately one third of subpath generations using an identical shooting point.

The region most strongly affected by repeated shooting-point selections corresponds to ensembles \( [5^+] \)--\( [7^+] \), with interfaces located in the \(0.449\)--\(0.504\)~nm range. This is also the region where a clear difference between the crossing-probability profiles has been established (Fig.~\ref{fig:main_results}(e)). However, the deviation between the curves begins at lower-index ensembles.
This can be understood from the sampling characteristics of conventional TIS. Low-index ensembles typically exhibit the highest degree of correlation because they sample relatively short paths at low potential energy.\cite{RETIS} Such short paths are more susceptible to becoming trapped in unfavorable channels that run parallel to the dominant reaction pathways along the order parameter. Consequently, low-index ensembles benefit strongly from replica-exchange moves. Paths reaching higher interfaces explore configurations further from the initial state and can potentially more readily cross barriers orthogonal to the order parameter.\cite{VANERP200834} Replica exchange can therefore transfer more diverse pathways back to the lower-index ensembles, enhancing their sampling and reducing correlations between successive paths.

Thus, although the shooting-point scarcity problem appears to become most pronounced from \( [5^+] \) onward, its effects can propagate to the lower-index ensembles. Limited exploration in the \( [5^+] \)--\( [7^+] \) region reduces the diversity of pathways available for replica exchange, thereby limiting the ability of these moves to introduce pathways that have explored orthogonal degrees of freedom into the lower-index ensembles.

Given the above statistics, which indicate that repeated shooting-point selections are most prevalent in the \( [5^+] \)--\( [7^+] \) ensembles, we further analyze this group as a whole in Table~\ref{tab2}. The top row reports the weighted average of the values for these three ensembles from Table~\ref{tab1}, where the weights are determined by the path count in each ensemble.
In the second row, we introduce the shooting-point diversity metric,
$D_{\mathrm{shot}}=N_{\mathrm{unique}}/N_{\mathrm{selected}}$. 

In the present case, each advanced shooting move consists of six successive subtrajectory generations, such that $D_{\mathrm{shot}}$ for each path-generation move ranges from $1/6$ to $6/6$. A value of $1/6$ corresponds to selecting the same shooting point in all six subtrajectory generations, whereas $6/6$ indicates that all six selected shooting points are unique. The average value of $D_{\mathrm{shot}}$ over all path-generation moves is shown in Table~\ref{tab2}. MRWF achieves an average shooting-point diversity close to 100\%, whereas the corresponding value for wire-fencing is only slightly above 50\%.
In parentheses, we also report the MRWF value obtained when restricting the analysis to low-resolution subpaths.  The shooting-point diversity is then even slightly higher,  indicating that the high-resolution subpaths do not artificially inflate the diversity of MRWF.

At first sight, a shooting-point diversity of slightly above 50\% for wire-fencing may not appear particularly low. However, this aggregate measure can mask the severity of the sampling problem because it averages over path-generation moves across different regions of the path ensemble. A high diversity in regions where sampling is easy can compensate for poor diversity in regions where the ability to explore alternative pathways is most important. In particular, the failure to generate diverse shooting points becomes most consequential when the system reaches configurations from which a new reaction pathway could potentially be discovered. 

To further substantiate our hypothesis that shooting-point selection is less diverse in wire-fencing than in MRWF, we considered three additional metrics. These quantify (i) the fraction of advanced shooting moves in which all selected shooting points are unique, (ii) the fraction of moves in which all subtrajectories are generated from exactly the same shooting point, corresponding to the situation depicted in Fig.~\ref{fig:oneshot}, and (iii) the fraction of moves in which three successive subtrajectories are generated from the same shooting point. For all three metrics, MRWF exhibits substantially greater shooting-point diversity than wire-fencing. The most striking difference is observed for the ``all the same shooting point'' metric: this situation never occurs for MRWF, whereas for wire-fencing it occurs in more than one quarter of all path-generation moves.

\begin{table}[!t]
\centering
\small
\caption{Shooting-point persistence and diversity in accepted wire-fencing and MRWF moves for ensembles $[5^+]$--$[7^+]$. The first row reports the fraction of successive shooting-point selections that returned to the same shooting point. The shooting-point diversity $D_{\mathrm{shot}}$ 
is calculated separately for each accepted path-generation move, and the table reports the mean over accepted moves.}
\label{tab2}
\begin{tabular}{lcc}
\hline
Diagnostic & WF & MRWF \tabularnewline
\hline
Consecutive selections repeated & 48.7\% & 3.63\% \tabularnewline
Mean $D_{\mathrm{shot}}$ & 57.4\% & 96.1\% (97.3\%) \tabularnewline
Moves with all points unique & 26.9\% & 82.0\% \tabularnewline
$N_{\rm unique}=1$ & 27.5\% & 0.0\% \tabularnewline
Moves with repeats $\geq 3$ & 54.3\% & 1.77\% \tabularnewline
\hline
\end{tabular}
\end{table}

\section{Conclusion}

We have developed the multi-resolution wire-fencing (MRWF) move, a new Monte Carlo move for path sampling designed to overcome shooting-point scarcity when trajectories are saved at a low frame-saving rate. By temporarily storing subpaths at higher resolution, MRWF increases the diversity of available shooting points without changing the output trajectory format, making it fully compatible with existing analysis tools.

We demonstrated that MRWF satisfies the required super-detailed balance conditions and converges to the same crossing probabilities, fluxes, and rate constants as conventional wire-fencing for two test systems. For the challenging T4L L99A/benzene ligand dissociation system, MRWF produced an approximately one-order-of-magnitude higher crossing probability, while reducing the estimated relative errors in the rate constant and crossing probability by factors of approximately 3.7 and 5.5, respectively. At the same computational cost, these improvements correspond to an estimated overall performance gain of approximately 14--30-fold relative to conventional wire-fencing. It is also worth noting that conventional wire-fencing has previously been shown to be substantially more efficient than standard shooting, which remains widely used as a default move in path-sampling studies.

Analysis of shooting-point diversity further indicates that MRWF provides access to a substantially larger variety of unique shooting points, supporting the hypothesis that it alleviates the sampling limitations caused by sparse trajectory saving. MRWF therefore represents a further step toward more efficient path sampling and is particularly promising for large biomolecular systems, where storage requirements can limit frame-saving rates, as well as for nucleation and other applications in which evaluation of the order parameter is computationally expensive.

\section*{Code availability}

The simulation input files for the double-well and puckering systems, together with the code used to generate and analyse the data presented in this work, are available through the \texttt{infRETIS} GitHub at \url{https://github.com/infretis}.

\begin{acknowledgments}
This work was supported by the Research Council of Norway through FRIPRO project 353364, COSY Gemini centre.
\end{acknowledgments}

\bibliographystyle{apsrev4-1}
\bibliography{biblio.bib}

@article{TIS,
	title = {A novel path sampling method for the calculation of rate constants},
	volume = {118},
	doi = {10.1063/1.1562614},
	number = {17},
	journal = {J. Chem. Phys.},
	author = {van Erp, Titus S. and Moroni, Daniele and Bolhuis, Peter G.},
	year = {2003},
	pages = {7762--7774},
}

@Article{RETIS,
  author  =  {T. {van Erp}},
  title   =  {Reaction rate calculation by parallel path swapping},
  journal =  {Phys. Rev. Lett.},
  year    =  {2007},
  volume  =  {98},
  pages   =  {268301}
}

@article{shoot,
	title = {Efficient transition path sampling: Application to Lennard-Jones cluster rearrangements},
	volume = {108},
	pages = {9236--9245},
	number = {22},
	journal = {J. Chem. Phys.},
	author = {Dellago, Christoph and Bolhuis, Peter G. and Chandler, David},
	year = {1998},
}

@article{PNAS2024,
  title        = {Highly parallelizable path sampling with minimal rejections using asynchronous replica exchange and infinite swaps},
  volume       = {121},
  pages        = {e2318731121},
  journal = {Proc. Natl. Acad. Sci. U. S. A.},
  author       = {Zhang, Daniel T. and Baldauf, Lukas and Roet, Sander and Lervik, Anders and van Erp, Titus S.},
  year         = {2024},
}

@article{WF,
  title        = {Enhanced path sampling using subtrajectory Monte Carlo moves},
  volume       = {158},
  pages        = {024113},
  journal = {J. Chem. Phys.},
  author       = {Zhang, Daniel T. and Riccardi, Enrico and van Erp, Titus S.},
  year         = {2023},
}

@article{moroni2004rate,
  title={Rate constants for diffusive processes by partial path sampling},
  author={Moroni, Daniele and Bolhuis, Peter G and van Erp, Titus S},
  journal={J. Chem. Phys.},
  volume={120},
  number={9},
  pages={4055--4065},
  year={2004},
  publisher={American Institute of Physics}
}

@article{faradjian2004computing,
  title={Computing time scales from reaction coordinates by milestoning},
  author={Faradjian, Anton K and Elber, Ron},
  journal={J. Chem. Phys.},
  volume={120},
  number={23},
  pages={10880--10889},
  year={2004},
}

@article{vervust2023path,
  title={Path sampling with memory reduction and replica exchange to reach long permeation timescales},
  author={Vervust, Wouter and Zhang, Daniel T and Van Erp, Titus S and Ghysels, An},
  journal={Biophys. J.},
  volume={122},
  number={14},
  pages={2960--2972},
  year={2023},
}

@article{cerou2011multiple,
  title={A multiple replica approach to simulate reactive trajectories},
  author={C{\'e}rou, Fr{\'e}d{\'e}ric and Guyader, Arnaud and Lelievre, Tony and Pommier, David},
  journal={J. Chem. Phys.},
  volume={134},
  number={5},
  year={2011},
}

@article{huber1996weighted,
  title={Weighted-ensemble Brownian dynamics simulations for protein association reactions},
  author = {Huber, Gary A. and Kim, Sangtae},
  journal={Biophys. J.},
  volume={70},
  number={1},
  pages={97--110},
  year={1996},
  publisher={Elsevier}
}

@article{InfRET1,
  title        = {Exchanging Replicas with Unequal Cost, Infinitely and Permanently},
  volume       = {126},
  pages        = {8878--8886},
  journal = {J. Phys. Chem. A},
  author       = {Roet, Sander and Zhang, Daniel T. and van Erp, Titus S.},
  year         = {2022},
}

@article{Rogal2010,
  author    = {J. Rogal and W. Lechner and J. Juraszek and B. Ensing and P. G. Bolhuis},
  title     = {The reweighted path ensemble},
  journal   = {J. Chem. Phys.},
  year      = {2010},
  volume    = {133},
  pages     = {174109},
}

@article{Hall2022RETISFFS,
  author  = {Hall, Steven W. and Díaz Leines, Grisell and Sarupria, Sapna and Rogal, Jutta},
  title   = {Practical guide to replica exchange transition interface sampling and forward flux sampling},
  journal = {J. Chem. Phys.},
  year    = {2022},
  volume  = {156},
  number  = {20},
  pages   = {200901},
  doi     = {10.1063/5.0080053}
}

@article{vanErp2016,
  author = {Titus S. van Erp and Mahmoud Moqadam and Enrico Riccardi and Anders Lervik},
  title = {Analyzing Complex Reaction Mechanisms Using Path Sampling},
  journal = {J. Chem. Theory Comput.},
  volume = {12},
  pages = {5398--5410},
  year = {2016},
}

@article{PyRETIS3,
  title = {{PyRETIS} 3: Conquering rare and slow events without boundaries},
  author = {Wouter Vervust and Daniel T. Zhang and An Ghysels and Sander Roet and Titus S. van Erp and Enrico Riccardi},
  journal = {J. Comput. Chem.},
  volume = {45},
  pages = {1224--1234},
  year = {2024}
}

@Article{TPSReview2,
  title =    {Transition path sampling and other advanced simulation techniques for rare events},
  journal =  "Adv. Poly. Sci. {\bf 221}, 167",
  author =   {C. Dellago and P.G. Bolhuis},
  year =     2009
}

@Article{TPS,
  title =    {Transition path sampling and the calculation of the rate constant},
  journal =  "J. Comput. Phys. {\bf 108}, 1964", 
  author =   {C. Dellago and P. G. Bolhuis and F. S. Csajka and D. Chandler},
  year =     1998
}

@misc{infretissoftware,
  author       = {{InfRETIS Development Team}},
  title        = {{InfRETIS: Python libraries for running $\infty$RETIS}},
  howpublished = {\url{https://github.com/infretis/}},
  year         = {2026},
  note         = {GitHub repository}
}

@article{vanErp2012,
  author    = {Titus S. van Erp},
  title     = {Dynamical Rare Event Simulation Techniques for Equilibrium and Nonequilibrium Systems},
  journal   = {Adv. Chem. Phys.},
  volume    = {151},
  pages     = {27--60},
  year      = {2012},
  doi       = {10.1002/9781118322182.ch2}
}

@article{riccardi2017fast,
author = {Riccardi, Enrico and Dahlen, Oda and van Erp, Titus S.},
title = {Fast Decorrelating Monte Carlo Moves for Efficient Path Sampling},
journal = {J. Phys. Chem. Lett.},
volume = {8},
number = {18},
pages = {4456-4460},
year = {2017},
doi = {10.1021/acs.jpclett.7b01617}
}

@article{vanErp2023RETIS,
  author  = {van Erp, Titus S.},
  title   = {How far can we stretch the timescale with RETIS?},
  journal = {EPL},
  year    = {2023},
  volume  = {143},
  number  = {3},
  pages   = {30001}
}

@book{Peters2017,
  author    = {Peters, Baron},
  title     = {Reaction Rate Theory and Rare Events},
  year      = {2017},
  publisher = {Elsevier}
}

@article{REW2026,
  title   = {Generalized path reweighting and history-dependent free energies},
  volume  = {164},
  number  = {21},
  pages   = {214107},
  journal = {J. Chem. Phys.},
  author  = {van Erp, Titus S. and Zhang, Daniel T. and Wils, Elias and Safaei, Sina and Ghysels, An},
  year    = {2026},
  doi     = {10.1063/5.0326023},
}

@article{Abraham2015GROMACS,
  author  = {Abraham, Mark James and Murtola, Teemu and Schulz, Roland and
             P{\'a}ll, Szil{\'a}rd and Smith, Jeremy C. and Hess, Berk and
             Lindahl, Erik},
  title   = {{GROMACS}: High performance molecular simulations through
             multi-level parallelism from laptops to supercomputers},
  journal = {SoftwareX},
  volume  = {1--2},
  pages   = {19--25},
  year    = {2015},
  doi     = {10.1016/j.softx.2015.06.001},
}

@article{Eriksson1992,
  author  = {Eriksson, A. E. and Baase, W. A. and Wozniak, J. A. and Matthews, B. W.},
  title   = {A cavity-containing mutant of T4 lysozyme is stabilized by buried benzene},
  journal = {Nature},
  volume  = {355},
  pages   = {371--373},
  year    = {1992},
  doi     = {10.1038/355371a0},
}

@article{Capelli2019,
  author  = {Capelli, Riccardo and Carloni, Paolo and Parrinello, Michele},
  title   = {Exhaustive Search of Ligand Binding Pathways via Volume-Based Metadynamics},
  journal = {J. Phys. Chem. Lett.},
  volume  = {10},
  number  = {12},
  pages   = {3495--3499},
  year    = {2019},
  doi     = {10.1021/acs.jpclett.9b01183},
}

@article{PLUMEDConsortium2019,
  author  = {{The PLUMED consortium}},
  title   = {Promoting transparency and reproducibility in enhanced molecular simulations},
  journal = {Nat. Methods},
  volume  = {16},
  number  = {8},
  pages   = {670--673},
  year    = {2019},
  doi     = {10.1038/s41592-019-0506-8},
}

@article{HajiAkbari2015,
  author  = {Haji-Akbari, Amir and Debenedetti, Pablo G.},
  title   = {Direct Calculation of Ice Homogeneous Nucleation Rate for a Molecular Model of Water},
  journal = {Proc. Natl. Acad. Sci. U.S.A.},
  volume  = {112},
  number  = {35},
  pages   = {10582--10588},
  year    = {2015},
  doi     = {10.1073/pnas.1509267112}
}

@article{Dietrich2024,
  author  = {Dietrich, Florian M. and Advincula, Xavier R. and Gobbo, Gianpaolo and Bellucci, Michael A. and Salvalaglio, Matteo},
  title   = {Machine Learning Nucleation Collective Variables with Graph Neural Networks},
  journal = {J. Chem. Theory Comput.},
  volume  = {20},
  pages   = {1600--1611},
  year    = {2024}
}

@article{Cheng2012,
  author  = {Y.-M. Cheng and S. M. Gopal and S. M. Law and M. Feig},
  title   = {Molecular Dynamics Trajectory Compression with a Coarse-Grained Model},
  journal = {IEEE/ACM Trans. Comput. Biol. Bioinform.},
  volume  = {9},
  number  = {2},
  pages   = {476--486},
  year    = {2012},
  doi     = {10.1109/TCBB.2011.141}
}

@article{Arjun2021,
  author  = {Arjun, A. and Bolhuis, P. G.},
  title   = {Molecular Understanding of Homogeneous Nucleation of CO2 Hydrates Using Transition Path Sampling},
  journal = {J. Phys. Chem. B},
  volume  = {125},
  pages   = {338--349},
  year    = {2021},
  doi     = {10.1021/acs.jpcb.0c09915}
}

@article{Feher1996,
author  = {Feher, Victoria A. and Baldwin, Enoch P. and Dahlquist, Frederick W.},
title   = {Access of ligands to cavities within the core of a protein is rapid},
journal = {Nature Structural Biology},
volume  = {3},
number  = {6},
pages   = {516--521},
year    = {1996},
doi     = {10.1038/nsb0696-516},
}

@article{Boothroyd2023Sage,
  author  = {Boothroyd, Simon and Behara, Pavan Kumar and Madin, Owen C. and Hahn, David F. and Jang, Hyesu and Gapsys, Vytautas and Wagner, Jeffrey R. and Horton, Joshua T. and Dotson, David L. and Thompson, Matthew W. and Maat, Jessica and Gokey, Trevor and Wang, Lee-Ping and Cole, Daniel J. and Gilson, Michael K. and Chodera, John D. and Bayly, Christopher I. and Shirts, Michael R. and Mobley, David L.},
  title   = {Development and Benchmarking of Open Force Field 2.0.0: The Sage Small Molecule Force Field},
  journal = {J. Chem. Theory Comput.},
  year    = {2023},
  volume  = {19},
  number  = {11},
  pages   = {3251--3275},
  doi     = {10.1021/acs.jctc.3c00039}
}

@misc{OpenFF210,
  author       = {Behara, Pavan Kumar and Gokey, Trevor and Cavender, Chapin and Horton, Joshua and Wang, Lily and Jang, Hyesu and Wagner, Jeffrey and Cole, Daniel J. and Bayly, Christopher I. and Mobley, David L.},
  title        = {{openforcefield/openff-forcefields}},
  year         = {2023},
  howpublished = {Zenodo},
  note         = {Version 2023.05.1},
  doi          = {10.5281/zenodo.7889050}
}

@article{Falkner2025,
  author  = {Falkner, Sebastian and Coretti, Alessandro and Peters, Baron and Bolhuis, Peter G. and Dellago, Christoph},
  title   = {Revisiting shooting point Monte Carlo methods for transition path sampling},
  journal = {J. Chem. Phys.},
  volume  = {163},
  number  = {3},
  pages   = {034105},
  year    = {2025},
  doi     = {10.1063/5.0261744}
}

@article{Ray2024,
  author  = {Ray, Dhiman and Parrinello, Michele},
  title   = {Data-driven classification of ligand unbinding pathways},
  journal = {Proceedings of the National Academy of Sciences},
  volume  = {121},
  number  = {10},
  pages   = {e2313542121},
  year    = {2024},
  doi     = {10.1073/pnas.2313542121}
}

@software{turtlemd,
  author  = {{infretis}},
  title   = {{TurtleMD}},
  year    = {2024},
  url     = {https://github.com/infretis/turtlemd},
  note    = {Slow molecular dynamics software for testing},
}

@article{VANERP200834,
title = {Efficient path sampling on multiple reaction channels},
journal = {Comput. Phys. Commun.},
volume = {179},
number = {1},
pages = {34-40},
year = {2008},
doi = {https://doi.org/10.1016/j.cpc.2008.01.023},
author = {Titus S. {van Erp}},
}

\end{document}